# Combinations of histone deacetylase inhibitors extend chronological lifespan in *S. cerevisiae*

Owen H. Wherry

**Aging is the primary risk factor for nearly all forms of human death, yet pharmaceutical interventions hold the potential to prevent it. Combinations of drugs have been shown to increase the lifespan of model organisms more than individual drugs, and geroprotective histone deacetylase (HDAC) inhibitors that have different molecular targets within the longevity regulation network show considerably higher drug synergy than many other compounds. In this study, four HDAC inhibitors (curcumin, quercetin, resveratrol, and berberine) have been administered in pairwise, three-, and four-combinations to yeast (*S. cerevisiae*) and their maximum chronological lifespans (CLS) have been measured. In five of the six pairwise combinations, the drugs exhibited synergy according to the Bliss Independence Model, on average extending maximum CLS 68% over the individual drugs. Three- and four-combinations further extended maximum CLS 49% and 107% over pairwise combinations, respectively. Since the targets of the HDAC inhibitors used in this study are evolutionarily conserved between yeast and humans, the results obtained have implications on human longevity.**

The biological process of aging accounts for over two-thirds of deaths globally, and as further improvements in early diagnostics, food and water, and hygiene reach developing areas, so will an increase in the number of chronic diseases in older populations (Li et al., 2023). As such, preventing the gradual physical, mental, emotional, and social decline of the human species has become a high-priority research endeavor (Moskalev et al., 2017). This approach starkly contrasts with the standard healthcare model, where human decline is only treated late in life when it manifests itself as aging-related diseases (Fig. 1).

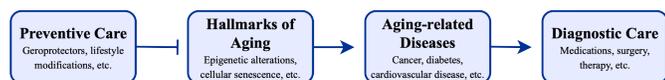

**Figure 1**. In a standard healthcare model, the hallmarks of aging slowly decrease the homeostatic capability of the human body until they manifest themselves as aging-related diseases and require treatment in the form of diagnostic care. In a preventive care model, the hallmarks of aging are targeted before they materialize as diseases, leading to lower healthcare costs and extending life in a period where the patient's quality of life is higher (Institute of Medicine of the National Academies, 2010).

However, out of the myriad of approaches available to increase the length of time that a person is alive (lifespan) and the length of time that a person is healthy (healthspan), few are as effective and practical as pharmaceutical interventions. While caloric restriction, reduced stress exposure, and other non-pharmaceutical interventions have been shown to lengthen lifespan and healthspan, they suffer from poor long-term adherence and multifaceted approaches that are hard to implement (Dorling et al., 2020; Fontana et al., 2010; Shields & Slavich, 2017). Geroprotectors, drugs that target the underlying hallmarks of aging in order to change the biological, functional, and digital data points that are predictors of mortality (biomarkers), however, are simple to implement, easy to adhere to, and cost-effective (Moskalev et al., 2017).

Yet the complexity of the longevity regulation network indicates that a single-drug approach is unlikely to significantly extend lifespan. It has been shown that over 50 unique genes contribute to longevity, and epigenetic factors, or how genes are expressed, also have a major impact and can be far more complex due to the diversity of our environments (Bin-Jumah et al., 2022). Synergistic combinations of drugs apt to target the complex molecular processes governing aging have shown promising results in many organisms, from plant extracts in *S. cerevisiae* to IGF-1 targeting geroprotectors in *C. elegans* to rapamycin and inhibitors of stress-activated pathways in *Drosophila melanogaster* and *Brachionus manjavacas* (Dakik et al., 2019; Admasu et al., 2018; Danilov et al., 2013; Snell et al., 2014)

Out of the over 200 compounds identified to have geroprotective effects and compiled into online databases like geroprotectors.org, epigenetic alteration mitigating histone deacetylase (HDAC) inhibitors have been shown to exhibit the highest rate of synergistic effects, particularly when each geroprotector affects a different node of the longevity regulation network (Moskalev et al., 2015; Rybina et al., 2023; Dakik et al., 2019).

This research focuses on four HDAC inhibitors with proven geroprotective effects, high consumer availability, and evidence of synergistic effects in other applications: curcumin, quercetin, resveratrol and berberine. Curcumin is an extract of *Curcuma longa* with anti-aging, anti-cancer, anti-hypertensive, anti-inflammatory, and anti-neurological effects (McCubrey et al., 2017). Quercetin is a flavonoid found in many plants (although at its highest concentration in capers) that has been shown to reduce inflammation, prevent mitochondrial dysfunction, and inhibit cancer growth among other effects (Rivero-Segura et al., 2024). Resveratrol is a polyphenol found in grape skins that shows anti-aging, anti-cancer, and anti-neurological properties in 110 clinical trials and in a plethora of environments and organisms (McCubrey et al., 2017). Berberine is an alkaloid found in *Berberis vulgaris* that demonstrates anti-inflammatory and anti-microbial properties and is frequently taken as an antidiabetic supplement (McCubrey et al., 2017).

In other applications, co-treatments of these drugs have shown synergistic effects: berberine and curcumin show synergistic chemoprotective effects on human breast cancer cells, berberine and resveratrol synergistically lower cholesterol levels in mice, and pairwise combinations of resveratrol, quercetin, and curcumin synergistically increase intestinal permeability (Wang et al., 2016; Zhu et al., 2018; Lund & Pantuso, 2014).

This study's objectives were (1) to measure the effects of pairwise and high-order combinations of curcumin, quercetin, resveratrol, and berberine on the chronological lifespan (CLS) of yeast and (2) to classify all six possible pairwise combinations of curcumin, quercetin, resveratrol, and berberine as synergistic or antagonistic and to state the degree to which they exhibit synergistic or antagonistic effects.

## Methods

To determine the effect of combinations of curcumin, quercetin, resveratrol, and berberine on lifespan, *S. cerevisiae* was used as a model organism and grown to the stationary phase prior to measurement. A high throughput CLS assay was chosen to take repeated, quantitative measurements of the cell viability via optical density readings, and then survival metrics were calculated to determine potential synergistic interactions between geroprotectors.

Although emergent synergistic reactions, or synergistic interactions that only occur in high-order combinations of drugs, can occur, they are very rare—representing only 3% of the information encapsulated by three-combinations—and the mean of pairwise interactions provides a sufficiently reliable estimate for synergy in three- or four-combinations (Cokol-Cakmak et al., 2020; Wood et al., 2012). As a result, three- and four-combinations were not classified as synergistic or antagonistic in this study and only their impact on CLS was measured along with their relative lifespan extension compared to the three constituent pairwise combinations or four constituent three-combinations for each three- and four-combination, respectively.

### Yeast strains, media, and growth conditions

The wild-type *Saccharomyces cerevisiae* (Carolina Biological Supply, Item #156250) was streaked onto a yeast extract/peptone dextrose (YPD) agar plate using a plastic disposable loop and placed in a 30°C incubator. YPD agar plates were made by dissolving 6.5% (w/v) of a premade YPD agar powder in deionized water, sterilizing the solution by autoclaving at 121°C for 15 minutes, cooling the solution in a 60°C water bath, and then pouring the solution into petri dishes and allowing time to dry overnight before storing upside down in a 4°C refrigerator. After the yeast was incubated for two days, a single colony was inoculated into the liquid medium. The experimental design for the growth conditions followed Dakik et al., where yeast was grown in 0.67% (w/v) synthetic minimal yeast nitrogen base (YNB) medium initially containing 2% glucose and supplemented with 20 mg/L histidine, 30 mg/L leucine, 30 mg/L lysine, and 20 mg/L uracil, giving yeast all essential nutrients for growth and reproduction (Dakik et al., 2019). Cells were then cultured at 30°C with rotational shaking at 200 rpm in 50 mL conical tubes with a volume/medium ratio of 5:1 to minimize the amount of evaporation (Parrella & Longo, 2008). After two days of culture, the cells reached the stationary phase and the first cell population was measured as part of the CLS assay (Wu et al., 2011). At all times when working with sterile media, a Bunsen burner set to a blue flame was used as well as sterile plastic pipette tips and inoculating loops, and all liquid media and agar was autoclaved at 121°C for 15 minutes prior to use to maintain sterility.

### Chronological lifespan assay

Due to the amenability of CLS measurement and time and equipment constraints, a high throughput assay developed by Wu et al. was used instead of replicative lifespan (RLS) measurements, more time-consuming assays such as the colony-forming unit (CFU) assay, or less precise assays like the semi-quantitative CLS assay using serial dilutions (Longo & Fabrizio, 2012; Wu et al., 2011; Powers et al., 2006; Arslan et al., 2018).

The experimental design followed published methods by Wu et al., where every 2-4 days 5.0 µL of the mixed culture was pipetted into each well of a 96-well microplate along with 100 µL of a YPD medium containing 1% yeast extract, 2% peptone, and 2% dextrose (Wu et al., 2011). The cell population was then calculated using a microplate reader by recording the optical density (OD) at a wavelength of 590 nm every 10 min for 24 hr to calculate the percentage of viable cells. This process was repeated every 2-4 days until the percentage of viable cells was calculated to be below 5%, at which point one more percentage was calculated. The percent survival of 5% was used to reduce the chances of a subpopulation utilizing the resources of dead cells to regrow, and it served as the maximum lifespan of the culture (Parrella & Longo, 2008).

### Histone deacetylase inhibitors

Berberine was added to liquid culture medium while quercetin, resveratrol, and curcumin were added to dimethyl sulfoxide (DMSO) such that each solute had a 20 mg/mL concentration and was fully dissolved. For cultures with two or more of curcumin, quercetin, and resveratrol, all solutes with the exception of berberine were prepared in the same DMSO solution to limit the final DMSO concentration to 5% (v/v). Additionally, a culture was prepared with a DMSO mock-solution to account for the possible impact of DMSO on yeast CLS despite evidence that concentrations of 5% DMSO or lower have no effect on yeast lifespan (Kakolyri et al., 2016). On the day of inoculating yeast into the liquid medium, quantities of the YNB medium, the berberine medium solution, and the quercetin, resveratrol, and curcumin DMSO solutions were added to obtain a 0.1% (w/v) final concentration of each HDAC inhibitor.

### Statistical analysis

Google Sheets was used to perform statistical analysis. The average doubling times of each set of absorbance curves were calculated and the cell viability percentage was obtained from that value (Fig. 2).

$$D_t = \frac{ln(2)}{\frac{ln(OD_2)-ln(OD_1)}{t_2 t_1}}; \quad V_n = \frac{1}{2^{\frac{\Delta t_n}{\overline{Dt_n}}}} * 100\%$$

**Figure 2**. Calculations of doubling time ($D_t$) and relative survival percentages ($V_n$) where $OD_1$ and $OD_2$ are two successive OD measurements, $t_1$ and $t_2$ are the times between measurements, $\overline{Dt_n}$ is the average doubling time defined as the average of the middle 50% of doubling times, and $n$ is in days (Wu et al., 2011).

To classify synergy among pairwise combinations, the Bliss Independence Model was determined to be the most appropriate (as opposed to the Highest Single Agent, Combination Subthresholding, or Response Additivity Models) since all of the geroprotectors target different signaling pathways—that is, they are mutually nonexclusive—so the value $E_A + E_B - E_A E_B$ was held as the requirement for synergy where $E_A$ is the effect of drug A, $E_B$ is the effect of drug B, and both

$E_A$ and $E_B$ are between 0 and 1 (Foucquier & Guedj, 2015). The combination index ($CI$) was then calculated as $CI = \frac{E_A + E_B - E_A E_B}{E_{AB}}$ where $E_{AB}$ is the effect of drug A and drug B in combination (Foucquier & Guedj, 2015). A $CI < 1$ indicates synergy while a $CI > 1$ indicates antagonism, with more extreme values representing greater synergy or antagonism, respectively.

**Results**

In isolation, each geroprotector extended maximum CLS to 6.0 days with the exception of berberine which extended maximum CLS to 8.5 days. This is significantly longer than the untreated maximum CLS of 3.0 days, where for the remainder of this research "untreated" will exclude data from the DMSO mock-treated culture since the maximum CLS values for the untreated and mock-treated cultures were statistically identical.

Pairwise combinations extended maximum CLS to 11.0 days, a 68% increase over single drug treatment. Five of the six combinations exhibited synergy while the combination of curcumin and berberine exhibited antagonism. Among the five cultures treated with pairwise combinations that exhibited synergy, an average maximum CLS of 11.3 days was observed, a 71% increase compared to single drug treatment.

Three- and four-combinations extended maximum CLS to 16.3 days and 22.8 days, respectively. This represents an additional 49% lifespan extension by the third drug administered and an additional 40% lifespan extension by the fourth drug administered.

***Pairwise combinations of curcumin and quercetin, curcumin and resveratrol, quercetin and resveratrol, quercetin and berberine, and resveratrol and berberine synergistically extend maximum yeast CLS***

The pairwise combination of curcumin and quercetin exhibited slight synergy with $CI = 0.99$ and extended maximum lifespan by an average of 3.5 days compared to the single drug cultures (Fig. 3).

The pairwise combination of curcumin and resveratrol exhibited synergy with $CI = 0.67$ and extended maximum lifespan by an average of 8.0 days compared to the single drug cultures (Fig. 4).

The pairwise combination of quercetin and resveratrol exhibited slight synergy with $CI = 0.99$ and extended maximum lifespan by an average of 3.5 days compared to the single drug cultures (Fig. 5).

The pairwise combination of quercetin and berberine exhibited slight synergy with $CI = 0.98$ and extended maximum lifespan by an average of 3.8 days compared to the single drug cultures (Fig. 6).

The pairwise combination of resveratrol and berberine exhibited synergy with $CI = 0.87$ and extended maximum lifespan by an average of 5.3 days compared to the single drug cultures (Fig. 7).

***A pairwise combination of curcumin and berberine does not synergistically extend maximum yeast CLS***

The pairwise combination of curcumin and berberine exhibited antagonism with $CI = 1.14$ and extended maximum lifespan by an average of 2.3 days compared to the single drug cultures (Fig. 8).

***Three- and four-combinations of berberine, quercetin, resveratrol, and curcumin significantly extended maximum yeast CLS compared to pairwise combinations***

The three-combination of curcumin, quercetin, and resveratrol extended maximum lifespan to 16.3 days, representing an average 47% increase over its constituent pairwise combinations (Fig. 9).

The three-combination of curcumin, quercetin, and berberine extended maximum lifespan to 20.0 days, representing an average 100% increase over its constituent pairwise combinations (Fig. 9).

The three-combination of curcumin, resveratrol, and berberine extended maximum lifespan to 16.3 days, representing an average 35% increase over its constituent pairwise combinations (Fig. 9).

The three-combination of quercetin, resveratrol, and berberine extended maximum lifespan to 12.5 days, representing an average 14% increase over its constituent pairwise combinations (Fig. 9).

The four-combination of curcumin, quercetin, resveratrol, and berberine extended maximum lifespan to 22.8 days, representing an average 40% lifespan extension over three-combinations (Fig. 9).

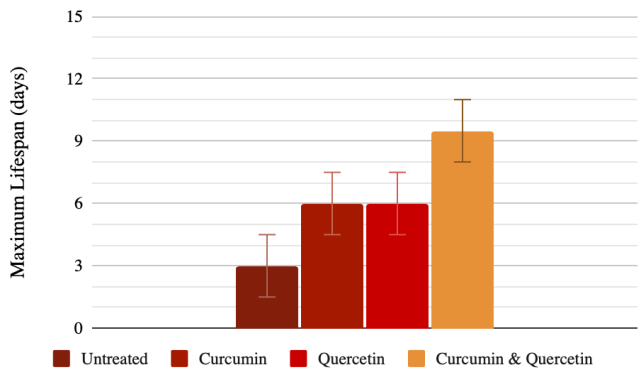

**Figure 3**. The combination of curcumin and quercetin extended maximum yeast CLS to 9.5 days compared to 6.0 or 3.0 days when one or no drugs were administered, respectively. With $CI = 0.99$, the combination of curcumin and quercetin is slightly synergistic.

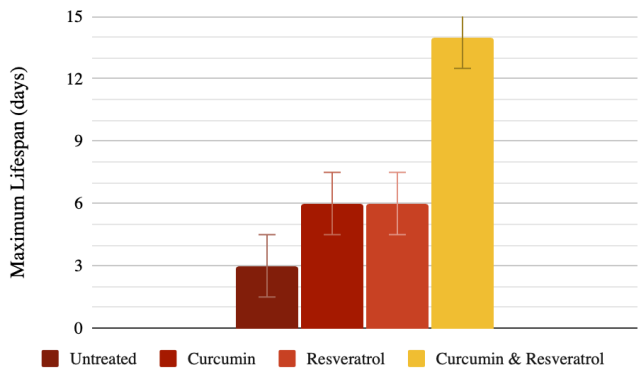

**Figure 4**. The combination of curcumin and resveratrol extended maximum yeast CLS to 14 days compared to 6.0 or 3.0 days when one or no drugs were administered, respectively. With $CI = 0.67$, the combination of curcumin and quercetin is highly synergistic.

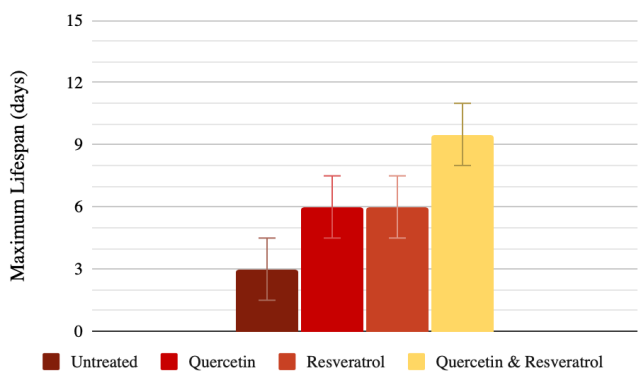

**Figure 5**. The combination of quercetin and resveratrol extended maximum yeast CLS to 9.5 days compared to 6.0 or 3.0 days when one or no drugs were administered, respectively. With $CI = 0.99$, the combination of quercetin and resveratrol is slightly synergistic.

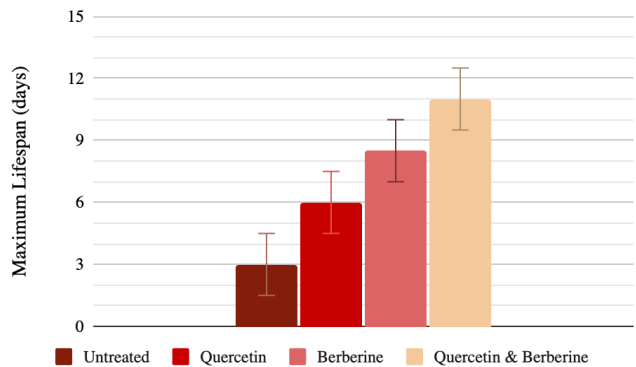

**Figure 6**. The combination of quercetin and berberine extended maximum yeast CLS to 11 days compared to 8.5, 6.0, or 3.0 days when only berberine, only quercetin, or no drugs were administered, respectively. With $CI = 0.98$, the combination of quercetin and berberine is slightly synergistic.

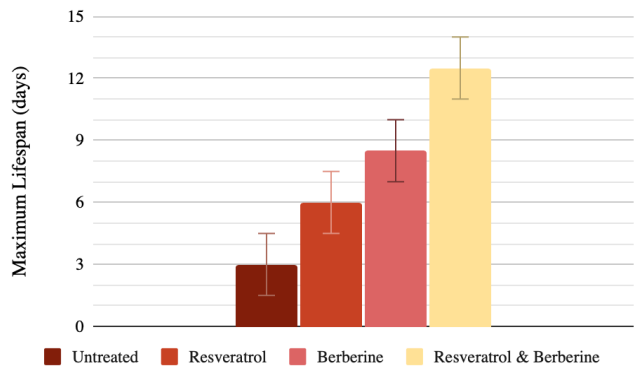

**Figure 7**. The combination of resveratrol and berberine extended maximum yeast CLS to 12.5 days compared to 8.5, 6.0, or 3.0 days when only berberine, only resveratrol, or no drugs were administered, respectively. With $CI = 0.87$, the combination of resveratrol and berberine is slightly synergistic.

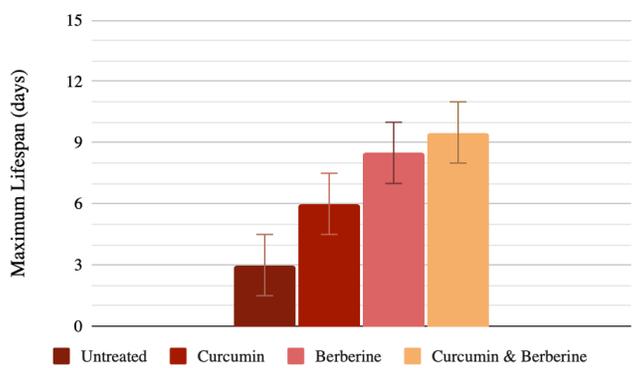

**Figure 8**. The combination of curcumin and berberine failed to extend lifespan significantly, resulting in a maximum CLS of 9.5 days compared to 8.5, 6.0, or 3.0 days when only berberine, only curcumin, or no drugs were administered, respectively. With $CI = 1.14$, the combination of curcumin and berberine is antagonistic.

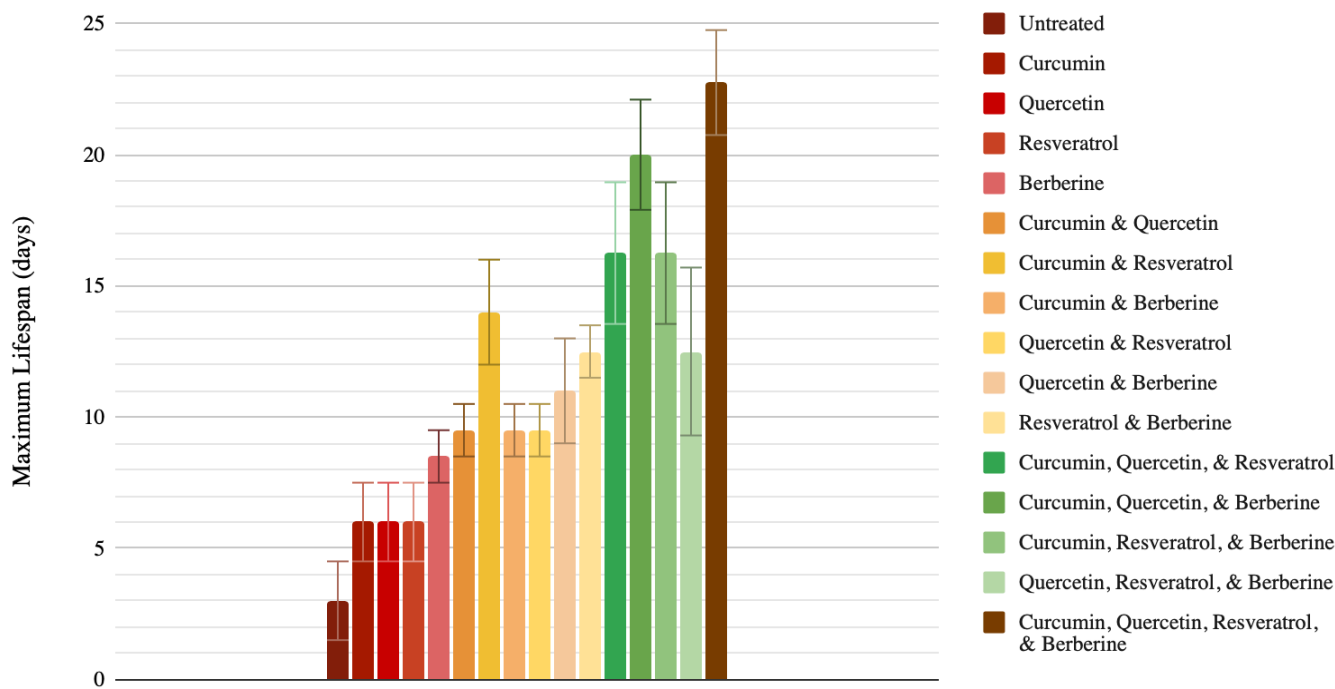

**Figure 9**. Maximum lifespans of yeast cultures treated with zero, one, two, three, or four geroprotective histone deacetylase inhibitors. On average, untreated yeast lived a maximum of 3.0 days while treatment with one, two, three, or four drugs extended maximum lifespan to 6.6, 11.0, 16.3, or 22.8 days, respectively.

## Discussion

With five out of six pairwise combinations showing some level of synergy, this study supports the idea that HDAC inhibiting geroprotectors exhibit high rates of synergy. Furthermore, the additional 49% or 40% maximum lifespan increase by administering a third or fourth drug, respectively, indicates that high-order combinations of synergistic drugs holds greater potential for significant lifespan extension than single drugs. It would be expected that these synergistic high-order drug combinations could be administered in lower doses, have fewer side effects, and target more aging pathways than single drugs (Lehár et al., 2009).

### *Limitations*

Several limitations were present in this study: mean lifespans of each culture were not measured, relatively infrequent measurements were taken, and equipment constraints were present.

While five of the six drug combinations tested synergistically extended maximum yeast CLS, other data points including mean lifespan, median lifespan, age of 90% mortality, and rate of aging were not measured and are helpful to determine the efficacy of individual geroprotectors or their combinations (Moskalev et al., 2016). Mean lifespan was attempted to be measured, but a combination of factors including differing starting absorbance values and a limited amount of total cultures (due to equipment constraints) yielded inconsistent and ultimately insignificant results. It is therefore possible that some combinations tested synergistically extend maximum lifespan without synergistically extending mean lifespan, a possibility that is present in combinations such as rapamycin and JNK SP600125 in *Brachionus manjavacas* or rifampicin and psora-4 in *C. elegans* (Snell et al., 2014; Admasu et al., 2018).

Additionally, the CLS assay was performed every 2-4 days for the duration of this study, resulting in a large amount of possible error among maximum lifespan measurements. This was particularly noticeable for the untreated culture due to its shorter maximum lifespan, however given that the untreated culture was primarily used to ensure viability and proper growth conditions the implications are largely insignificant.

Finally, limited space in shaking incubators mandated a smaller number of cultures with the same experimental conditions than is desired, increasing the potential for error, especially in three- and four-combinations.

*Implications*

Future research should focus on two primary goals: (1) screening for synergistic high-order combinations of geroprotectors, and (2) testing of combinations of curcumin, quercetin, resveratrol, and berberine in other organisms to determine efficacy in humans.

While high-order combinations can be tested for synergy *in vivo* such as in this study, it is more efficient to screen high-order combinations with other methods since the number of experiments scales exponentially with the number of drugs tested (Zimmer et al., 2016). Previously reliable methods to predict effective high-order combinations include stochastic search algorithms and machine learning models that use sets of pharmaceuticals, disease targets, and dose levels—however, the most effective models strongly base their predictions off of pairwise *in vivo* and/or *in vitro* measurements (Vakil & Trappe, 2019; Larkins-Ford et al., 2022). As a result, the data obtained in this study could be used as input data for machine learning algorithms to predict effective high-order combinations.

Furthermore, this research has implications on human longevity. Despite yeast's evolutionary divergence from humans and other animals around one billion years ago, yeast and humans still share over 4,300 genes and even more cellular pathways, including many of the essential pathways involved in nutrient signaling and other aging processes (Kachroo et al., 2022; Fontana et al., 2010). Therefore, drug combinations such as the ones explored in this study are likely to have similar effects in human cells—however, to ensure amply efficacy, rule out possible negative effects, and determine optimal dosages, additional testing in other model organisms and clinical trials are necessary before use in humans (Institute of Medicine of the National Academies, 2009).